\documentclass[11pt]{article}

\usepackage[utf8]{inputenc}
\usepackage[T1]{fontenc}
\usepackage{helvet} 

\usepackage{graphicx}
\usepackage[margin=1in]{geometry}
\usepackage[dvipsnames]{xcolor}
\definecolor{softblue}{HTML}{34568B}
\definecolor{slategray}{HTML}{708090}
\usepackage{hyperref}
\hypersetup{
    colorlinks=true,
    citecolor=softblue,      
    urlcolor=softblue,       
    linkcolor=slategray,     
    filecolor=slategray,      
}
\usepackage{parskip}

\usepackage{booktabs}
\usepackage{longtable}
\usepackage{array}
\usepackage{rotating}
\usepackage{tabularx}
\usepackage{adjustbox}
\usepackage{capt-of}
\usepackage{float}

\newcolumntype{P}[1]{>{\raggedright\arraybackslash}p{#1}}

\usepackage{amsmath}
\usepackage{listings}

\lstdefinestyle{javastyle}{
  language=Java,
  backgroundcolor=\color{white},   
  commentstyle=\color{gray},
  keywordstyle=\color{blue},
  numberstyle=\tiny\color{gray},
  stringstyle=\color{purple},
  basicstyle=\ttfamily\footnotesize,
  breakatwhitespace=false,         
  breaklines=true,                 
  captionpos=b,                    
  keepspaces=true,                 
  numbers=left,                    
  numbersep=5pt,                  
  showspaces=false,                
  showstringspaces=false,
  showtabs=false,                  
  tabsize=2
}

\title{Assessing the Quality and Security of AI-Generated Code: A Quantitative Analysis}
\author{
  \begin{tabular}{c}
    Abbas Sabra, Olivier Schmitt, Joseph Tyler \\
    \\
    Sonar
  \end{tabular}
}
\date{\today}

\begin{document}

\maketitle

\begin{abstract}
This study presents a quantitative evaluation of the code quality and security of five prominent Large Language Models (LLMs): Claude Sonnet 4, Claude 3.7 Sonnet, GPT-4o, Llama 3.2 90B, and OpenCoder 8B. While prior research has assessed the functional performance of LLM-generated code \cite{chen}, this research tested LLM output from 4,442 Java coding assignments through comprehensive static analysis using SonarQube \cite{sonarqube}. The findings suggest that although LLMs can generate functional code, they also introduce a range of software defects, including bugs, security vulnerabilities, and code smells. These defects do not appear to be isolated; rather, they may represent shared weaknesses stemming from systemic limitations within current LLM code generation methods. In particular, critically severe issues, such as hard-coded passwords \cite{cwe798, owaspAuth} and path traversal vulnerabilities \cite{cwe22, owaspAccess}, were observed across multiple models. These results indicate that LLM-generated code requires verification in order to be considered production-ready. This study found no direct correlation between a model's functional performance (measured by \textsc{Pass@1} rate of unit tests) and the overall quality and security of its generated code, measured by the number of SonarQube issues in benchmark solutions that passed the functional tests. This suggests that functional benchmark performance score is not a good indicator of overall code quality and security. The goal of this study is not to rank LLM performance but to highlight that all evaluated models appear to share certain weaknesses. Consequently, these findings support the view that static analysis can be a valuable instrument for detecting latent defects and an important safeguard for organizations that deploy AI in software development.
\end{abstract}

\newpage
\tableofcontents
\newpage

\section{Introduction}
State of the art LLMs from providers such as Anthropic \cite{anthropic}, OpenAI \cite{openai}, and Meta \cite{meta} are increasingly utilized in software development for various tasks, including code generation, code autocompletion, and bug remediation. The potential for these tools to enhance developer productivity and accelerate development drives their widespread adoption. However, accelerated development must be balanced with the foundational principles of software quality. Core attributes like reliability, security, and maintainability are crucial for the long-term viability of any software project. Code that fails to meet these standards, whether authored by humans or generated by AI, can accrue technical debt \cite{martini2015, behutiye2024}, introduce security risks, and degrade system reliability.

The prevailing discourse on LLMs has largely emphasized their generating capabilities, such as measuring functional correctness \cite{chen}, sometimes overshadowing the critical need to evaluate the broader quality and security of their output \cite{Dora, fan2023}. This concern is amplified by the rapid adoption of LLMs in software development, with reports indicating AI assistants write an average of 46\% of developer code \cite{zhao2023github}, while research continues to highlight that the generated code can introduce significant security vulnerabilities and bugs \cite{Dora, fan2023}. Consequently, there is an established risk of developers overlooking necessary verification, as studies have found that programmers using AI assistants can produce less secure code while simultaneously showing greater confidence in its security \cite{Dora}.

\newpage
\section{Research Objectives}
This investigation seeks to address the aforementioned quality gap. The study is guided by the following central research questions (RQs), which focus on identifying common patterns in the output of leading LLMs rather than on providing a comparative ranking:

\begin{enumerate}
    \item[RQ1:] What is the typical characteristic quality profile of Java code generated by contemporary, state-of-the-art LLMs?
    \item[RQ2:] What are the most common categories and severity levels of issues observed in the code generated by these models?
    \item[RQ3:] How effectively can static analysis identify these defects and serve as a protective mechanism?
    \item[RQ4:] Does an improvement in a model's functional performance correlate with an improvement in the quality and security of its generated code?
\end{enumerate}

This research investigates the gap between the output of LLMs and the quality standards for production-ready software. We posit that LLM-generated code is not immediately fit for production and requires rigorous verification, making functional metrics alone insufficient for evaluation. This study argues that the common error patterns in LLM-generated code make static analysis an essential tool for ensuring its quality and security.

\newpage
\section{Experiment Setup}

This section outlines the experiment design used to evaluate the code quality and security of several leading LLMs using a set of Java programming challenges. The structure facilitates a cross-model comparison and the application of static analysis to identify issue patterns in the generated code.

\subsection{Benchmarks}

The evaluation was conducted using Java as the target programming language. The benchmark dataset comprised 4,442 distinct coding problems from three recognized and publicly available sources: \textsc{MultiPL-E-mbpp-java}, \textsc{MultiPL-E-humaneval-java} \cite{cassano2023}, and \textsc{ComplexCodeEval} \cite{feng2024}. While code was generated for static analysis from all three sources, functional performance (i.e., test pass rates) was evaluated using only the two \textsc{MultiPL-E} benchmarks. These benchmarks were selected to represent a spectrum of Java programming challenges with varying complexity and domain relevance, forming a robust testbed for assessing the LLMs' code generation capabilities.

\subsection{Models Under Test}

The study assessed five prominent LLMs, representing diverse architectures, training methods, and accessibility:
\begin{itemize}
    \item \textbf{Claude Sonnet 4 and Claude 3.7 Sonnet (Anthropic):} Included to examine how code generation quality evolves between successive model generations from the same provider \cite{anthropic}.
    \item \textbf{GPT-4o (OpenAI):} Selected for its widespread adoption, as it frequently serves as the underlying technology for popular AI coding tools such as GitHub Copilot \cite{openai}.
    \item \textbf{Llama 3.2 90B (Meta):} Chosen as a leading open-weight model representative of state-of-the-art, openly available LLMs \cite{meta}.
    \item \textbf{OpenCoder-8B (Llama-3.1-8B architecture):} A smaller, open-source model included to determine if common code quality issues persist regardless of model size, which could reveal challenges inherent in the LLM methodology itself \cite{huang2024}.
\end{itemize}
For consistency and reproducibility, identical prompts were provided to each model, and the temperature was set to zero.

\subsection{Methodology: Code Generation and Static Analysis}

Our methodology involved a three-phase process:
\begin{itemize}
    \item \textbf{Code Generation:} Each of the five LLMs was prompted to generate compilable Java solutions for all 4,442 benchmark problems. The primary objective was to obtain functional and compilable code to serve as the input for the subsequent quality assessment.
    \item \textbf{Functional Performance Evaluation:} The code generated for problems from the two \textsc{MultiPL-E} benchmarks was executed against the provided test suites to measure functional performance. This performance was quantified by \textsc{Pass@1} rate of unit tests.
    \item \textbf{Cross-Model Static Analysis:} Each generated Java file was then analyzed using SonarQube \cite{sonarqube}. The full default SonarWay Java ruleset, consisting of approximately 550 rules designed to be broadly suitable for most projects, was applied to detect various categories of software issues, including bugs, security vulnerabilities, and code smells. This allowed us to compare the distribution and prevalence of issues across models, identify shared weaknesses, and characterize the quality of AI-generated code.
\end{itemize}
This design enabled a detailed examination of not only whether LLMs could solve coding problems but also how well they did so from a quality and security perspective.

\newpage
\section{Quality Overview of LLM-Generated Code}

This section addresses the characteristic quality profile of LLM-generated Java (RQ1) by providing a quantitative overview of the code generated for the 4,442 tasks. The analysis focuses on overarching characteristics and aggregated quality metrics.

\subsection{Volumetric and structural characteristics}

The five LLMs exhibited notable differences in the volume and complexity of the Java code they generated. These metrics, detailed in Table \ref{tab:codeMetrics}, provide context for the subsequent analysis of code quality.

\begin{table}[H]
\centering
\caption{Comparative code generation metrics across LLMs (4,442 Tasks)}
\label{tab:codeMetrics}
\resizebox{\textwidth}{!}{%
\begin{tabular}{lccccccccc}
\toprule
\textbf{LLM Model} & \textbf{LOC} & \textbf{Statements} & \textbf{Functions} & \textbf{Classes} & \textbf{Files} & \textbf{Comment Lines} & \textbf{Comments (\%)} & \textbf{Cyclomatic Complexity} & \textbf{Cognitive Complexity} \\
\midrule
Claude Sonnet 4 & 370,816 & 148,932 & 46,235 & 12,832 & 4,442 & 20,051 & 5.10\% & 81,667 & 47,649 \\
Claude 3.7 Sonnet & 288,126 & 116,433 & 27,496 & 10,649 & 4,442 & 56,459 & 16.40\% & 55,485 & 42,220 \\
GPT-4o & 209,994 & 83,466 & 24,309 & 10,475 & 4,442 & 9,692 & 4.40\% & 44,387 & 26,450 \\
Llama 3.2 90B & 196,927 & 75,368 & 22,694 & 8,996 & 4,442 & 15,514 & 7.30\% & 37,948 & 20,811 \\
OpenCoder-8B & 120,288 & 41,510 & 8,338 & 5,530 & 4,442 & 13,165 & 9.90\% & 18,850 & 13,965 \\
\bottomrule
\end{tabular}%
}
\end{table}

These metrics reveal distinct generative tendencies. Claude Sonnet 4 produced the most code (370,816 LOC) with the highest cumulative Cyclomatic Complexity (81,667) \cite{mccabe1976} and Cognitive Complexity (47,649) \cite{campbell2018}. In contrast, OpenCoder-8B generated the most concise code (120,288 LOC) with the lowest complexity scores (18,850 Cyclomatic Complexity and 13,965 Cognitive Complexity). Comment density also varied widely, from Claude 3.7 Sonnet’s high of 16.4\% to GPT-4o’s 4.4\% and Claude Sonnet 4’s 5.1\%.

The variance in structural metrics suggests that the choice of an LLM has significant consequences for a project's long-term maintainability, irrespective of the model's functional performance. This finding indicates that different architectural philosophies and training data result in varied structural outputs, rather than a consistent progression toward more concise or less complex code in newer models. This diversity underscores a key conclusion: LLM-generated code is not a monolithic entity, and development teams should adopt model selection and code review strategies that account for these structural differences.

\subsection{Functional Performance and Overall Code Quality}

Beyond structural characteristics, a central question of this study is whether a model’s functional performance—defined here as the success rate of generated code passing its intended tests—correlates with the quality of its generated code (RQ4). This study measured functional performance using the test pass rates from the \textsc{MultiPL-E} benchmarks, while static analysis was conducted on the code generated for all 4,442 tasks. Table \ref{tab:passRates} summarizes these metrics to provide an initial overview of code quality.

\begin{table}[H]
\centering
\caption{LLM Performance: Test Pass Rates and SonarQube Issue Rates (4,442 Tasks)}
\label{tab:passRates}
\resizebox{0.9\textwidth}{!}{%
\begin{tabular}{lccc}
\toprule
\textbf{LLM Model} & \textbf{Passing tests \%} & \textbf{SonarQube Discovered Issues} & \textbf{Issues per passing task} \\
\midrule
Claude Sonnet 4 & 77.04 & 7,225 & 2.11 \\
Claude 3.7 Sonnet & 72.46 & 6,576 & 2.04 \\
GPT-4o & 69.67 & 5,476 & 1.77 \\
Llama 3.2 90B & 61.47 & 5,159 & 1.89 \\
OpenCoder-8B & 60.43 & 3,903 & 1.45 \\
\bottomrule
\end{tabular}%
}
\end{table}

Claude Sonnet 4 demonstrated the highest test pass rate at 77.04\%, while OpenCoder-8B exhibited the lowest at 60.43\%. The "Issues per Passing Task" metric, which normalizes total issues against the number of functionally successful outputs, offers additional insight. OpenCoder-8B, despite its lower pass rate, presented the lowest number of issues per passing task (1.45). In contrast, Claude Sonnet 4, the top performer on pass rate, averaged 2.11 issues per passing task.

A key observation is that even when LLM-generated code passes functional performance benchmarks, it is not free of underlying quality defects. For every task that OpenCoder-8B completed successfully, it still averaged 1.45 static analysis issues; similarly, GPT-4o averaged 1.77 issues per passing task. This pattern supports the premise that static code analysis is valuable for assessing the quality of functionally "passing" code. These underlying issues in successful tasks represent latent factors that could impact maintainability and reliability over time.

The consistent presence of quality issues in functionally correct code suggests that relying solely on functional performance benchmarks to evaluate LLM-generated code is insufficient and may introduce hidden risks. This highlights a potential paradox: as models become more capable, they may generate more sophisticated solutions that, while functionally robust, introduce a larger surface area for defects, leading to a greater number of static analysis findings. This dynamic is examined more closely in the following sections.

\subsection{Overall issue density and distribution by type and severity}

This section examines the types and severities of issues identified in the Java code generated by the LLMs across the 4,442 tasks. Table \ref{tab:defectMetrics} provides an overview of the resulting defect metrics.

\begin{table}[H]
\centering
\caption{Overall code quality defect metrics per LLM (4,442 Tasks)}
\label{tab:defectMetrics}
\resizebox{\textwidth}{!}{%
\begin{tabular}{lcccc}
\toprule
\textbf{LLM Model} & \textbf{SonarQube Discovered Issues} & \textbf{LOC} & \textbf{Issues per KLOC} & \textbf{Issues per Passing Task} \\
\midrule
Claude Sonnet 4 & 7,225 & 370,816 & 19.48 & 2.11 \\
Claude 3.7 Sonnet & 6,576 & 288,126 & 22.82 & 2.04 \\
GPT-4o & 5,476 & 209,994 & 26.08 & 1.77 \\
Llama 3.2 90B & 5,159 & 196,927 & 26.20 & 1.89 \\
OpenCoder-8B & 3,903 & 120,288 & 32.45 & 1.45 \\
\bottomrule
\end{tabular}%
}
\end{table}

Issue density, measured as issues per thousand lines of code (KLOC), ranged from 19.48 for Claude Sonnet 4 to 32.45 for OpenCoder-8B. These differing densities appear to illustrate distinct profiles of issue generation rather than a simple quality ranking. A model that generates a larger volume of code might exhibit a higher total number of issues, even if its per-line quality is comparatively reasonable. The distribution of issues by type, shown in Table \ref{tab:issueTypes}, reveals similarity across models.

\begin{table}[H]
\centering
\caption{Distribution of issue types by LLM (absolute counts and percentage of total issues per model)}
\label{tab:issueTypes}
\resizebox{\textwidth}{!}{%
\begin{tabular}{lcccccc}
\toprule
\textbf{LLM Model} & \textbf{Total Bugs} & \textbf{\% Bugs} & \textbf{Total Vulnerabilities} & \textbf{\% Vulnerabilities} & \textbf{Total Code Smells} & \textbf{\% Code Smells} \\
\midrule
Claude Sonnet 4 & 423 & 5.85\% & 141 & 1.95\% & 6,661 & 92.19\% \\
Claude 3.7 Sonnet & 352 & 5.35\% & 116 & 1.76\% & 6,108 & 92.88\% \\
GPT-4o & 406 & 7.41\% & 112 & 2.05\% & 4,958 & 90.54\% \\
Llama 3.2 90B & 398 & 7.71\% & 123 & 2.38\% & 4,638 & 89.90\% \\
OpenCoder-8B & 247 & 6.33\% & 67 & 1.72\% & 3,589 & 91.95\% \\
\bottomrule
\end{tabular}%
}
\end{table}

A key observation from the data is the striking similarity in the distribution of issue types across all evaluated models. Despite their varied architectures, each model produced a comparable mix of defects: approximately 90-93\% code smells, 5-8\% bugs, and around 2\% security vulnerabilities (as shown in Table \ref{tab:issueTypes}). This consistency across different models suggests a systemic pattern in the code generation process of current LLMs. While code smells are the most frequent, the consistent introduction of bugs and, more critically, security vulnerabilities is particularly noteworthy. A defect rate that includes a 5-8\% chance of being a bug and a $\approx$2\% chance of being a security vulnerability is significant. This finding underscores that LLM-generated code, even when it passes functional performance tests, is not immediately suitable for production environments. It highlights the critical need for rigorous static analysis and expert human review to identify and remediate these underlying quality and security issues before deployment, thereby preventing the accumulation of technical debt and security risks. Table \ref{tab:densityByType} details the issue density by type for each model.

\begin{table}[H]
\centering
\caption{Issue density by type and LLM (per KLOC)}
\label{tab:densityByType}
\resizebox{0.9\textwidth}{!}{%
\begin{tabular}{lccc}
\toprule
\textbf{LLM Model} & \textbf{Bug Density (Bugs/KLOC)} & \textbf{Vulnerability Density (Vuln./KLOC)} & \textbf{Code Smell Density (Smells/KLOC)} \\
\midrule
Claude Sonnet 4 & 1.14 & 0.38 & 17.96 \\
Claude 3.7 Sonnet & 1.22 & 0.40 & 21.20 \\
GPT-4o & 1.93 & 0.53 & 23.61 \\
Llama 3.2 90B & 2.02 & 0.62 & 23.55 \\
OpenCoder-8B & 2.05 & 0.56 & 29.84 \\
\bottomrule
\end{tabular}%
}
\end{table}

Despite variations in density, all evaluated models produced all three types of issues: bugs, vulnerabilities, and code smells.

\subsection{Issue severity distribution (\% of Total Issues)}

This section analyzes the proportional distribution of severity levels (Blocker, Critical, Major, Minor) within each issue type to provide insight into the potential impact of the defects generated by the models. Table \ref{tab:bugDist} shows the severity distribution for bugs.

\begin{table}[H]
\centering
\caption{Bug distribution (\% of total bugs per model)}
\label{tab:bugDist}
\resizebox{0.7\textwidth}{!}{%
\begin{tabular}{lcccc}
\toprule
\textbf{LLM Model} & \textbf{BLOCKER} & \textbf{CRITICAL} & \textbf{MAJOR} & \textbf{MINOR} \\
\midrule
Claude Sonnet 4 & 13.71 & 4.49 & 54.14 & 27.66 \\
Claude 3.7 Sonnet & 7.10 & 3.98 & 61.93 & 26.99 \\
GPT-4o & 7.14 & 3.45 & 74.63 & 14.78 \\
Llama 3.2 90B & 13.82 & 4.77 & 56.78 & 24.62 \\
OpenCoder-8B & 9.24 & 12.05 & 49.00 & 29.72 \\
\bottomrule
\end{tabular}%
}
\end{table}

In the bug category, GPT-4o exhibited a notable tendency, with nearly 75\% of its bugs categorized as `MAJOR', indicating a propensity to produce significant functional defects. In contrast, Claude Sonnet 4 and Llama 3.2 90B presented the highest proportion of `BLOCKER' bugs (approximately 14\%), which represent defects that can prevent application functionality. OpenCoder-8B showed a high percentage of `CRITICAL' bugs (12\%) compared to other models, which generally ranged between 3\% and 5\%. For code smells, the severity distribution is presented in Table \ref{tab:smellDist}.

\begin{table}[H]
\centering
\caption{Code smell distribution (\% of total code smells per model)}
\label{tab:smellDist}
\resizebox{0.7\textwidth}{!}{%
\begin{tabular}{lcccc}
\toprule
\textbf{LLM Model} & \textbf{BLOCKER} & \textbf{CRITICAL} & \textbf{MAJOR} & \textbf{MINOR} \\
\midrule
Claude Sonnet 4 & 0.25 & 8.89 & 51.22 & 39.65 \\
Claude 3.7 Sonnet & 0.30 & 12.96 & 45.65 & 41.09 \\
GPT-4o & 0.22 & 5.79 & 43.45 & 50.54 \\
Llama 3.2 90B & 0.33 & 5.36 & 39.64 & 54.67 \\
OpenCoder-8B & 0.31 & 5.49 & 37.28 & 56.93 \\
\bottomrule
\end{tabular}%
}
\end{table}

For code smells, which relate primarily to maintainability, most issues were distributed across the `MAJOR' and `MINOR' categories for all models. Claude 3.7 Sonnet generated a markedly higher proportion of `CRITICAL' code smells (13\%) compared to the other models. Conversely, Llama 3.2 90B and OpenCoder-8B exhibited the highest percentages of `MINOR' code smells, suggesting their maintainability issues were, on average, of less severe immediate impact. The severity distribution for vulnerabilities, shown in Table \ref{tab:vulnDist}, highlights a more critical trend.

\begin{table}[H]
\centering
\caption{Vulnerability distribution (\% of total vulnerabilities per model)}
\label{tab:vulnDist}
\resizebox{0.7\textwidth}{!}{%
\begin{tabular}{lcccc}
\toprule
\textbf{LLM Model} & \textbf{BLOCKER} & \textbf{CRITICAL} & \textbf{MAJOR} & \textbf{MINOR} \\
\midrule
Claude Sonnet 4 & 59.57 & 28.37 & 5.67 & 6.38 \\
Claude 3.7 Sonnet & 56.03 & 28.45 & 5.17 & 10.34 \\
GPT-4o & 62.50 & 23.21 & 5.36 & 8.93 \\
Llama 3.2 90B & 70.73 & 22.76 & 1.63 & 4.88 \\
OpenCoder-8B & 64.18 & 26.87 & 1.49 & 7.46 \\
\bottomrule
\end{tabular}%
}
\end{table}

An important finding from the analysis of vulnerabilities is that all models produced a high percentage of `BLOCKER' and `CRITICAL' vulnerabilities. This observation underscores the importance of thorough security scanning for AI-generated code. For example, Llama 3.2 90B generated a high proportion of these issues, with over 70\% of its identified vulnerabilities classified as `BLOCKER'. Similarly, results for OpenCoder-8B and GPT-4o indicated that nearly two-thirds of their detected vulnerabilities were of the highest severity levels.
\newpage
\section{Analysis of Code Smells in LLM-Generated Java}
\label{sec:code_smells}

In addressing the most common categories of issues (RQ2), this section provides a detailed analysis of code smells. Code smells can serve as indicators of deeper structural problems within source code. While not representing direct functional errors, they can impede maintainability, comprehensibility, and evolvability, often contributing to the accumulation of technical debt \cite{martini2015, behutiye2024} or the introduction of bugs over time. As noted in the previous section, code smells \cite{fowler1999} were the most frequent issue type identified across all evaluated LLMs. This section provides an analysis of the specific sub-categories of code smells observed, as detailed in Table \ref{tab:smellCategories}, and explores potential factors that may contribute to these challenges for LLMs. For additional details regarding the categories, please refer to the Appendix.

\begin{table}[H]
\centering
\caption{Sub-categories of code smells and their origins (\% of total code smells for model)}
\label{tab:smellCategories}
\resizebox{\textwidth}{!}{%
\begin{tabular}{l c c c c c l}
\toprule
\textbf{Category} & \textbf{\shortstack{Claude Sonnet\\4 (\%)}} & \textbf{\shortstack{Claude 3.7\\Sonnet (\%)}} & \textbf{\shortstack{GPT-4o\\(\%)}} & \textbf{\shortstack{Llama 3.2\\90B (\%)}} & \textbf{\shortstack{OpenCoder-8B\\(\%)}} & \textbf{Potential Contributing Factors} \\
\midrule
Dead / Unused / Redundant code & 14.83 & 17.43 & 26.3 & 34.82 & 42.74 & Requires non-local, project-wide reference analysis. \\
Design / Framework best-practices & 22.26 & 18.58 & 20.81 & 18.84 & 12.45 & Lacks context of project-specific framework conventions. \\
Assignment / Field / Scope visibility & 11.96 & 15.35 & 13.21 & 11.32 & 11.95 & Requires class-wide or non-local scope resolution. \\
Collection / Generics / Param / Type & 13.94 & 11.23 & 9.92 & 9.03 & 7.89 & Requires deep semantic understanding of the API. \\
Regex / Pattern / String / Format & 13.70 & 11.80 & 7.36 & 6.81 & 5.29 & Logical flaws may only be apparent at runtime. \\
Cognitive / Computational complexity & 4.25 & 8.43 & 3.73 & 2.67 & 2.79 & Complexity is a non-local property of the code. \\
Control / Conditional-logic smell & 4.67 & 3.91 & 4.03 & 3.02 & 2.20 & Difficulty in balancing correctness and readability. \\
Deprecated / Obsolete APIs & 2.01 & 2.34 & 2.08 & 2.89 & 4.01 & Requires knowledge of library deprecation roadmaps. \\
Naming / Style / Documentation & 2.69 & 2.50 & 2.84 & 2.16 & 1.89 & Difficulty generalizing from project-specific conventions. \\
Exception-handling smell & 0.05 & 0.08 & 0.06 & 0.02 & 0.06 & Requires analysis of the cross-package dependency graph. \\
Other & 9.64 & 8.33 & 9.64 & 8.41 & 8.72 & \\
\bottomrule
\end{tabular}%
}
\end{table}

The following analysis delves into the most prominent categories from Table \ref{tab:smellCategories}, exploring potential factors that contribute to them.

\textbf{Dead / Unused / Redundant Code:} This category was prevalent in the output of models such as Llama 3.2 90B (34.82\% of its code smells) and OpenCoder-8B (42.74\%). LLMs may struggle in this area, potentially because identifying such code can require a whole-project reference analysis. Since LLMs often operate with a limited context window, it can be challenging for them to determine if a generated element is utilized elsewhere in a larger application. This limitation could lead to the generation of syntactically plausible but unreferenced code, contributing to codebase bloat.

\textbf{Design / Framework Best Practices:} Claude Sonnet 4 showed a higher percentage of issues in this category (22.26\%), which may reflect the model's tendency to generate more thorough code by attempting to handle numerous edge cases. This approach, while sometimes leading to more robust error management, can also create unnecessary logical complexity if the model cannot infer the specific context. However, all LLMs may face a common challenge in this area, as adherence to specific design patterns or framework conventions can depend on knowledge of organizational rules or proprietary mechanisms that may not be well-represented in general training data.

\textbf{Assignment / Field / Scope Visibility:} This issue was consistently observed, typically ranging from 11\% to 15\% of total code smells per model. This type of issue may arise because correctly determining scope can require a comprehensive, class-wide context. Defining the narrowest necessary scope for a variable might depend on an understanding of all interactions within a class, which could be challenging for a model to infer from localized generation.

\textbf{Collection / Generics / Parameter / Type Issues:} Issues in this category may be common because the correct use of generics and parameterized types often requires a deeper understanding of API semantics. While LLMs may employ these features accurately in simple cases, ensuring type safety across complex interactions could require a level of semantic comprehension that may be challenging for current generative models.

\textbf{Regex / Pattern / String / Format:} LLMs may generate regular expressions or string formatting operations that are syntactically valid but contain subtle logical flaws. This type of issue may occur because such flaws are often revealed only through execution or deeper semantic analysis, which is typically outside the scope of token-by-token generation.

\textbf{Cognitive / Computational Complexity:} High cognitive complexity was a notable issue, particularly for Claude 3.7 Sonnet (8.43\% of its code smells). LLMs may face challenges with this metric, as complexity is often a non-local property of code. Since LLMs tend to optimize for local token probability, a sequence of individually plausible segments might cumulatively result in a method that is globally complex, as the model may not have an explicit mechanism to optimize for an overall complexity score.

\textbf{Control / Conditional-Logic Smell:} LLMs may generate convoluted conditional logic. This issue may arise because capturing the nuance between functional correctness and code readability can be difficult. A model might not prioritize simpler or more idiomatic control flow structures if a more complex alternative also appears to satisfy the immediate generation objective.

\textbf{Deprecated / Obsolete APIs:} The use of deprecated APIs and outdated dependencies poses a significant security risk, moving beyond a simple code smell. An LLM's tendency to favor older API versions—a behavior likely influenced by its training data's knowledge cut-off or the prevalence of older code examples—can inadvertently lead to the re-introduction of libraries with known vulnerabilities (CVEs). This specific blind spot highlights the necessity of complementing static analysis with Software Composition Analysis (SCA). While static analysis inspects the generated code itself, SCA provides the crucial, additional security layer of scanning its dependencies for known exploits, directly addressing a risk vector that is particularly pronounced in how LLMs currently operate.

\textbf{Naming / Style / Documentation:} While LLMs may adhere to common coding conventions, they may not capture team- or project-specific naming styles. Such conventions vary widely across contexts, making them difficult for a model to generalize from disparate training data.

\textbf{Exception-Handling Smell:} The use of generic exceptions appears to be a frequent code smell across the evaluated models. Formulating specific exception handling can require a detailed analysis of dependencies across a codebase, which is a challenge for common use of LLMs that operate primarily within a localized context.
\newpage
\section{Analysis of Bugs in LLM-Generated Java}
\label{sec:bugs}

Continuing the investigation into the most common categories and severity levels of defects (RQ2), this section examines the bugs identified in the LLM-generated Java code. Bugs represent functional defects in code that can lead to incorrect behavior, application crashes, or unexpected outcomes. Static analysis detects bugs that can potentially compromise application stability and correctness. While numerically less frequent than code smells, their potential impact is often more immediate and severe. This section explores common bug categories observed in the LLM-generated Java code, which are summarized in Table \ref{tab:bugCategories}, and discusses potential factors contributing to their introduction.

\begin{table}[H]
\centering
\caption{Sub-categories of bugs and their origins (\% of total bugs for model)}
\label{tab:bugCategories}
\resizebox{\textwidth}{!}{%
\begin{tabular}{l c c c c c l}
\toprule
\textbf{Category} & \textbf{\shortstack{Claude Sonnet\\4 (\%)}} & \textbf{\shortstack{Claude 3.7\\Sonnet (\%)}} & \textbf{\shortstack{GPT-4o\\(\%)}} & \textbf{\shortstack{Llama 3.2\\90B (\%)}} & \textbf{\shortstack{OpenCoder-8B\\(\%)}} & \textbf{Potential Contributing Factors} \\
\midrule
Control-flow mistake & 14.83 & 23.62 & 48.15 & 31.06 & 21.37 & Requires deep, non-local path reasoning beyond pattern matching. \\
API contract violation & 10.29 & 14.12 & 8.64 & 14.90 & 19.35 & Requires analysis of error propagation across multiple code branches. \\
Exception handling & 16.75 & 16.71 & 11.60 & 14.39 & 14.52 & Dependent on understanding library intent and return semantics. \\
Resource management / Leak & 15.07 & 8.36 & 7.41 & 12.88 & 9.68 & Resource lifecycle management is a non-local problem. \\
Type-safety / Casts & 11.24 & 12.97 & 7.90 & 6.82 & 7.66 & Requires precise tracking of static-type provenance. \\
Concurrency / Threading & 9.81 & 1.44 & 1.73 & 1.26 & 2.82 & Concurrency concepts (e.g., atomicity) are underrepresented in corpora. \\
Null / Data-value issues & 7.89 & 7.49 & 8.89 & 5.81 & 6.85 & Difficulty tracking nullability across complex data-flow paths. \\
Performance / Structure & 4.31 & 6.34 & 3.95 & 2.78 & 5.24 & Potential generation of inefficient or suboptimal algorithms. \\
Pattern / Regex & 2.63 & 1.15 & 0.74 & 0.25 & 2.42 & Regex logical errors are often only evident at runtime. \\
Data-structure bug & 1.44 & 1.15 & 0.00 & 1.01 & 1.61 & Proper collection usage is contingent on semantic intent. \\
Serialization / Serializable & 0.00 & 0.58 & 0.00 & 0.76 & 1.61 & Requires knowledge of the framework's object-graph semantics. \\
Other & 5.74 & 6.05 & 0.99 & 8.08 & 6.85 & \\
\bottomrule
\end{tabular}%
}
\end{table}

A closer look at these categories, presented in Table \ref{tab:bugCategories}, reveals common challenges LLMs face in ensuring the semantic correctness and runtime integrity of generated code.

\textbf{Control-Flow Mistakes:} This category of bugs was particularly prominent in code from GPT-4o (48.15\% of its total bugs) and was also significantly represented in output from Llama 3.2 90B (31.06\%) and OpenCoder-8B (21.37\%). These issues may be prevalent because ensuring correct control flow can require deep path reasoning. While models can generate plausible conditional statements or loops based on learned patterns, they may face challenges with the multi-step logic needed to ensure correctness across all execution paths, especially with intricate branching or edge cases.

\textbf{API Contract Violations:} Models from the Llama family, specifically OpenCoder-8B (19.35\%) and Llama 3.2 90B (14.90\%), showed a higher proportion of issues in this area. Correctly using Application Programming Interfaces (APIs) can require analyzing error propagation and understanding return value semantics. LLMs might overlook these nuances, potentially leading to bugs where API return values are ignored or misinterpreted. This could suggest a challenge for the models in comprehending sequential, stateful operations.

\textbf{Exception Handling (Bugs):} Distinct from exception handling smells, this category pertains to functional defects. Addressing them can require knowledge of library intent. LLMs might generate code that catches overly broad exceptions without reacting appropriately or that disregards checked exceptions, potentially due to challenges in inferring the specific purpose of library-thrown exceptions from training data.

\textbf{Resource Management / Leaks:} Claude Sonnet 4 demonstrated a higher proportion of issues in this category (15.07\%). Resource leaks may be a persistent issue, possibly because resource lifecycles (e.g., opening and closing streams) can span multiple calls, which may be difficult to track within a local context window. This could lead to failures in ensuring that resources are properly closed across all paths, potentially introducing risks for long-running applications.

\textbf{Type-Safety / Casts:} Issues such as illegal type casts may occur because ensuring cast safety can require precise tracking of a variable's static-type history. A model might generate code involving type casting without being able to fully track the type history through complex data flows, which could result in runtime ClassCastException errors.

\textbf{Concurrency / Threading:} Claude Sonnet 4 also showed a comparatively higher share of bugs in this category (9.81\%). Correct concurrent programming is known to be difficult, and LLMs may face challenges in this area because concepts like thread state and atomicity can be complex to learn. Models may not always generate code that sufficiently covers complex threading scenarios, which could lead to potential race conditions or deadlocks.

\textbf{Null / Data-Value Issues:} This was a common problem, accounting for 5-9\% of identified bugs across the evaluated models. NullPointerExceptions may arise if a model faces difficulties tracking nullability across complex data flows. As a result, it might not consistently perform necessary null checks before dereferencing an object.

\textbf{Performance / Structure (Bugs):} LLMs may generate code that is functionally correct but exhibits poor runtime performance (e.g., inefficient loops). Such issues might occur because LLMs tend to optimize for local token generation probability rather than for global performance characteristics.

\textbf{Pattern / Regex (Bugs):} This category involves regular expressions that are functionally incorrect. These issues may arise because regex edge cases often surface only during execution or deeper semantic analysis, which may extend beyond the focus on syntactic validity during generation.

\textbf{Data-Structure Bugs:} The misuse of data structures, such as attempting to access an element beyond array bounds, may occur because the correct application of collections can be tied to semantic intent, a nuance that might be challenging for an LLM to capture fully.

\textbf{Serialization / Serializable:} Bugs such as a class failing to implement Serializable when required may stem from the non-local, framework-dependent nature of such requirements, which can be challenging for context-limited generation. The nature of these observed bugs suggests that LLMs may face challenges with the semantic correctness and runtime implications of the code they generate, potentially prioritizing syntactic plausibility over comprehensive functional integrity.
\newpage
\section{Analysis of Security Vulnerabilities in LLM-Generated Java}
\label{sec:vulnerabilities}

As a critical component of understanding the common issue categories (RQ2), this section analyzes the security vulnerabilities found in the generated code. Security vulnerabilities represent exploitable flaws in code that can lead to data compromise, service disruption, or unauthorized access. While comprising a smaller percentage of total issues (approximately 2\% across models), the 67 distinct types of vulnerabilities identified in this analysis are of significant concern due to their potential impact. Recent studies have highlighted these risks, identifying vulnerabilities in code generated by a range of models \cite{Dora}, which aligns with general industry awareness of key security risks \cite{owaspTop10}. A potential contributing factor to the prevalence of these vulnerabilities is that LLMs optimize for token likelihood based on training data, which may include insecure or outdated code snippets. Table \ref{tab:vulnCategories} provides a detailed breakdown of the vulnerability sub-categories identified in this study.

\begin{table}[H]
\centering
\caption{Sub-categories of security vulnerabilities and their origins (\% of total vulnerabilities for model)}
\label{tab:vulnCategories}
\resizebox{\textwidth}{!}{%
\begin{tabular}{l c c c c c l}
\toprule
\textbf{Category} & \textbf{\shortstack{Claude Sonnet\\4 (\%)}} & \textbf{\shortstack{Claude 3.7\\Sonnet (\%)}} & \textbf{\shortstack{GPT-4o\\(\%)}} & \textbf{\shortstack{Llama 3.2\\90B (\%)}} & \textbf{\shortstack{OpenCoder-8B\\(\%)}} & \textbf{Potential Contributing Factors} \\
\midrule
Path-traversal \& Injection & 34.04 & 31.03 & 33.93 & 26.83 & 28.36 & Requires non-local taint analysis from a data source to a sink. \\
Hard-coded credentials & 14.18 & 10.34 & 17.86 & 23.58 & 29.85 & Security-sensitive intent of constants is semantically unclear. \\
Cryptography misconfiguration & 24.82 & 23.28 & 19.64 & 22.76 & 22.39 & Requires current knowledge of secure cryptographic algorithms and modes. \\
XML External Entity (XXE) & 10.64 & 15.52 & 13.39 & 19.51 & 5.97 & Secure parser configuration is a non-local, multi-file problem. \\
Inadequate I/O error-handling & 4.96 & 7.76 & 7.14 & 4.88 & 7.46 & Semantic distinction between critical and benign errors is subtle. \\
Certificate-validation omissions & 2.84 & 4.31 & 2.68 & 0.00 & 2.99 & Secure, library-specific SSL/TLS usage patterns are rare in corpora. \\
JSON-injection risk & 0.71 & 0.00 & 0.89 & 0.81 & 1.49 & Trust boundaries are often implicit in data builders and serializers. \\
JWT signature not verified & 0.00 & 0.00 & 0.00 & 0.00 & 1.49 & Requires current, library-specific security token best practices. \\
Other & 7.80 & 7.76 & 4.46 & 1.63 & 0.00 & \\
\bottomrule
\end{tabular}%
}
\end{table}

Analysis of the vulnerabilities detailed in Table \ref{tab:vulnCategories} points to systemic weaknesses related to non-local taint analysis and the semantic understanding of sensitive data.

\textbf{Path-Traversal \& Injection:} This category was a dominant type of vulnerability across all models, observed, for instance, in 34.04\% of Claude Sonnet 4's vulnerabilities and 33.93\% of GPT-4o's. Preventing such flaws can require taint-tracking from an input source to a sensitive sink, a form of non-local analysis. The models may generate code that performs a function correctly but does not fully account for how unvalidated user input could manipulate file paths or inject commands. This challenge in performing comprehensive data flow analysis is a known security concern \cite{cwe22, owaspAccess}.

\textbf{Hard-Coded Credentials:} This critical vulnerability was particularly prevalent in Llama-family models, appearing in 29.85\% of OpenCoder-8B's vulnerabilities and 23.58\% of Llama 3.2 90B's. This may arise because constant strings can appear benign, and their security-sensitive intent may not be apparent from common patterns in training data. A password string literal, for example, might not be treated as semantically distinct from any other string, particularly if similar insecure practices are present in the training corpus \cite{cwe798, owaspAuth}.

\textbf{Cryptography Misconfiguration:} This was another significant area of weakness, exemplified by Claude Sonnet 4 at 24.82\% of its vulnerabilities. Secure cryptography often requires precise knowledge of secure versus weak cipher algorithms and modes. LLMs may reproduce patterns involving deprecated or weak cryptographic primitives, possibly because their training data contains such examples and may not reflect the most up-to-date standards.

\textbf{XML External Entity (XXE):} Vulnerabilities related to XXE may occur because correct configurations of XML parsers can span multiple files. Lacking full application context of how the parser is configured, an LLM might generate XML parsing code with insecure default settings \cite{cwe611, owaspMisconfig}.

\textbf{Inadequate I/O Error-Handling (Security):} Failures to properly handle I/O errors can lead to security issues. LLMs may face challenges in this area, as distinguishing between security-critical and routine operational errors can be a subtle distinction that requires deeper contextual understanding.

\textbf{Certificate-Validation Omissions:} Secure SSL/TLS communication typically requires proper certificate validation. LLMs may omit these steps, possibly because comprehensive, library-specific SSL/TLS usage patterns are underrepresented or oversimplified in general training data.

\textbf{JSON-Injection Risk:} LLMs may fail to respect the trust boundaries implicit in JSON builders or serializers, resulting in insecure constructions when unsanitized user-controlled data is incorporated.

\textbf{JWT Signature Not Verified:} Secure handling of JSON Web Tokens (JWT) generally requires verification of the token's signature. An LLM might omit this step, possibly because doing so requires adherence to up-to-date, library-specific best practices that may not be current or comprehensively represented in its training data.

The consistency of these vulnerabilities across different LLMs points to their training data as the most plausible origin. By learning from and replicating insecure code, the models demonstrate an inability to distinguish secure patterns from insecure ones, revealing a critical limitation in the current code generation paradigm. Therefore, while prompt engineering is useful, it is not a sufficient safeguard on its own. To mitigate these risks, applying rigorous security-focused static analysis (SAST) and expert human review to LLM-generated code is essential, particularly when it handles sensitive data or operations.
\newpage
\section{Discussion: Examples of Common SonarQube Detected Issues}

This section provides concrete examples of common SonarQube detections, building upon the detailed categorization of Code Smells, Bugs, and Vulnerabilities presented in Sections \ref{sec:code_smells}, \ref{sec:bugs}, and \ref{sec:vulnerabilities}. These examples are intended to illustrate potential shared weaknesses in LLM-generated code. The focus is on how specific detections may relate to the broader issue categories discussed in those sections, rather than on model-specific prevalence.

\subsection{Theme: Deficient Error Handling}

\textbf{Illustrative Rule:} \texttt{java:S112} -- Define and throw a dedicated exception instead of using a generic one.

\textbf{Observation:} A common pattern observed across the evaluated LLMs was the use of generic exceptions (e.g., \texttt{throw new Exception()}) rather than specific, custom exceptions. This was frequently flagged by SonarQube rule \texttt{java:S112} and was among the top issues for models like Claude Sonnet 4, GPT-4o, and Llama 3.2 90B.

\textbf{Interpretation:} This observation may align with the challenge related to exception handling, as discussed in Section \ref{sec:code_smells}, which can require a "cross-package dependency graph analysis." Generating specific exceptions often involves a broader understanding of an application's error handling strategy and class hierarchy—a level of non-local context that may be challenging for an LLM to infer. This behavior is also symptomatic of a tendency of LLMs to lack specificity[add ref?], perhaps in order to avoid hallucinations; the generated code will avoid execution errors, but at the cost of understanding the nuance of the scenario the exception is designed to handle.

\subsection{Theme: Resource Management Lapses}

\textbf{Illustrative Rule:} \texttt{java:S2095} -- Use try-with-resources or close this resource in a "finally" clause.

\textbf{Observation:} Failure to properly close resources, such as streams or network connections, was a recurring Blocker-level bug. Rule \texttt{java:S2095} identified numerous instances of this across different models, including 54 instances for Claude Sonnet 4, 25 for GPT-4o, and 50 for Llama 3.2 90B.

\textbf{Interpretation:} This bug category may be related to the challenge where a resource life-cycle spans many calls, making it difficult to manage. An LLM might lack the reasoning capability to consistently plan and track which resources are opened, and therefore fail to ensure its closure on all possible execution paths.

\subsection{Theme: Critical Security Oversights}

\textbf{Illustrative Rule:} \texttt{java:S6437} -- Revoke and change this password, as it is compromised - i.e., hardcoded password \cite{cwe798}.

\textbf{Observation:} The Blocker vulnerability of hardcoded credentials was detected in code generated by all five evaluated LLMs. For instance, 20 instances were found for Claude Sonnet 4, 20 for GPT-4o, and 29 for Llama 3.2 90B.

\textbf{Interpretation:} This vulnerability arises from the model's indiscriminate handling of string literals. The LLM does not differentiate between security-sensitive constants, such as passwords or API keys, and benign string values. Consequently, it embeds sensitive data directly into the source code as hardcoded constants. This behavior is consistent with a model designed to replicate plausible statistical patterns observed in its training data, which often includes insecure examples, rather than performing security-specific semantic analysis. 

\subsection{Theme: Excessive Code Complexity}

\textbf{Illustrative Rule:} \texttt{java:S3776} -- Refactor this method to reduce its Cognitive Complexity.

\textbf{Observation:} Many LLMs, notably Claude 3.7 Sonnet (422 instances) and GPT-4o (112 instances), generated methods with high Cognitive Complexity, which was flagged as a Critical issue by SonarQube.

\textbf{Interpretation:} Autoregressive LLMs generate tokens one at a time and are optimized to ensure generated code is locally coherent. Code architecture is of secondary importance during training, and at inference time the accumulating complexity of a given method is not tracked, therefore it is not surprising that LLMs generate code with high complexity, especially for tasks which require more sophisticated architecture.

\subsection{Theme: Maintainability \& Best Practice Violations (Redundant Code)}

\textbf{Illustrative Rule:} \texttt{java:S2094} -- Remove this empty class, write its code or make it an "interface".

\textbf{Observation:} The generation of empty classes or methods was a frequent Minor code smell across multiple models, including GPT-4o (531 instances) and OpenCoder-8B (661 instances).

\textbf{Interpretation:} This appears to be a manifestation of the "Dead / Unused / Redundant code" smell category. This may be linked to the challenge of performing a "whole-project reference analysis, not snippet." An LLM might generate such placeholder structures as part of a broader pattern but not subsequently populate them or identify their redundancy. These examples collectively suggest that common SonarQube detections can be indicative of potential systemic weaknesses in current LLM code generation approaches. The consistency of these anti-patterns across diverse LLMs may suggest that they are not merely random errors, but result from patterns learned from training data, or from challenges in translating high-level requirements into robust and maintainable code. These thematic groupings suggest that LLMs may face challenges with aspects of software engineering that involve foresight, strategic planning, and an understanding of non-local consequences—qualities often associated with the engineering discipline in software development. As we saw for error handing, this could be another example of LLMs avoiding hallucinations by lacking specificity, in this case generating pseudo-implementations.
\newpage
\section{Case Study: Evolution of Model Performance and Defect Characteristics}

To directly address whether an improvement in a model's functional performance correlates with an improvement in code quality (RQ4), this section presents a case study comparing two models from different generations. The analysis between Claude 3.7 Sonnet and its successor, Claude Sonnet 4, indicates that while benchmark performance can improve, the nature and severity of certain flaws might increase.

\subsection{Performance Benchmarks and Defect Severity}

\textbf{Improved Functional Performance:} The newer model, Claude Sonnet 4, achieved a higher benchmark score by passing 77.04\% of tasks, compared to the older model's 72.46\%. This suggests progress in the model's primary function of generating code that passes given tests.

\textbf{Increased Bug Severity:} An analysis of the generated bugs indicates a notable trend. The proportion of `BLOCKER' bugs nearly doubled in Claude Sonnet 4, increasing to 13.71\% from Claude 3.7 Sonnet's 7.1\%.

\textbf{Increased Vulnerability Severity:} A similar pattern was observed for security vulnerabilities. The proportion of `BLOCKER' vulnerabilities rose from 56.03\% in the older model to 59.57\% in the newer model, suggesting that when the newer model introduced a vulnerability, it had a higher probability of being of the highest severity.

\subsection{Persistence and Evolution of Underlying Issues}

\textbf{Persistent Code Quality Issues:} The general profile of issues in the generated code appears largely consistent across model versions. For both models, code smells constituted the majority of flaws (over 92\%). Despite its higher benchmark score, the newer Claude 4 still generated 2.11 issues for every test it passed, which may indicate that improved functional performance does not necessarily equate to a corresponding improvement in all code quality attributes.

\textbf{Evolution of the Defect Profile:} Comparing the model versions reveals an evolving defect profile, not a simple reduction of flaws. Although the general defect categories persisted, their distribution shifted—for instance, the newer Claude 4 model produced a higher proportion of `Concurrency / Threading' bugs. This suggests that changes in training strategies cause fundamental code generation challenges to evolve rather than be resolved.

This case study suggests that progress in AI models is not uniform across all quality attributes. An increase in benchmark scores may paradoxically accompany more severe bugs and vulnerabilities, which underscores the critical role of rigorous static code analysis when applying LLMs to software development.
\newpage
\section{The Role of Static Analysis in Addressing Potential LLM-Generated Issues}

The findings from the preceding sections regarding systemic issues in LLM generated code lead to the questions of detection and mitigation. To that end, this section addresses how effectively static analysis can serve as a protective mechanism for development teams using AI assisted tools (RQ3). In this capacity, static analysis, as exemplified by SonarQube in this study, offers a valuable safeguard by providing an automated and consistent mechanism for flagging known anti-patterns, including resource management lapses and critical security vulnerabilities, before they enter a codebase.

The value of a tool like SonarQube can be seen in its capacity to detect specific, and often critical, flaws of the types observed in the LLM-generated code. For instance, the security vulnerability of hardcoded credentials (SonarQube rule \texttt{java:S6437}) \cite{cwe798} was consistently observed in this study. As discussed in Section \ref{sec:vulnerabilities}, this appears to be a common pitfall for LLMs, possibly related to challenges in discerning the sensitive nature of certain constants. Similarly, resource management lapses, such as unclosed streams (flagged by \texttt{java:S2095}), were identified as recurring Blocker-level bugs. This finding may align with the previously noted challenge for LLMs in managing resource lifecycles that extend beyond a local context view.

Furthermore, the code smell of high Cognitive Complexity (flagged by \texttt{java:S3776}) was common in the output of several LLMs, including Claude 3.7 Sonnet (422 instances) and GPT-4o (112 instances). This may reflect a tendency for LLMs to optimize for local token generation, potentially without accounting for global complexity metrics. Such complex code can impede maintainability and testability.

In these examples, static analysis tools like SonarQube can provide an automated and consistent mechanism for flagging known anti-patterns before they creep into a codebase. The comprehensive rule sets of such tools are often designed to cover a wide spectrum of issues—including bugs, vulnerabilities, and code smells—which aligns well with the types of potential flaws identified in this study. Automated detection may be particularly valuable because manual review, while important, might not consistently identify all such issues, especially when managing large volumes of code.

Static analysis tools may become particularly important within an LLM-driven development paradigm, as they can provide a consistent, objective baseline for quality and security that may not be an inherent feature of probabilistic generative models. As LLMs can introduce variability, the deterministic and rule-based nature of static analysis may offer a valuable safeguard.

As LLMs become more integral to software development, the function of static analysis tools could evolve from a quality and security assurance measure to a component of responsible AI adoption. Static analysis may help bridge the gap between the output of LLMs and the quality and security standards often required in professional software engineering. The integration of such tools into Continuous Integration/Continuous Deployment (CI/CD) pipelines can further enhance the value of static analysis by allowing for continuous validation of code contributions.

\newpage
\section{Conclusion}

This quantitative analysis of five prominent LLMs over 4,442 Java tasks provides clear insights into the quality and security of AI-generated code. The findings address our four primary research questions, leading to a central conclusion: LLMs are powerful but imperfect coding assistants, and their output must be rigorously verified.

First, the study confirms that all evaluated models produce issues. None generated consistently defect-free code, instead introducing a diverse spectrum of defects (RQ1). We classify these defects into the three analyzed types: code smells that degrade maintainability (e.g., excessive complexity), bugs that impact runtime reliability (e.g., resource leaks), and critical security vulnerabilities (RQ2). Security vulnerabilities include severe flaws such as hardcoded passwords \cite{cwe798, owaspAuth}, path traversal vulnerabilities \cite{cwe22, owaspAccess}, and XML External Entity (XXE) injection flaws \cite{cwe611, owaspMisconfig}, demonstrating that static analysis is an effective mechanism for identifying these latent risks (RQ3).

Second, the research shows no correlation between a model's functional performance and the quality of its code (RQ4). This leads to several critical insights for model selection. We found that bigger is not necessarily better, as a model's scale or novelty did not guarantee higher-quality output. Critically, smaller can be just as good or better, with smaller models sometimes producing cleaner code for the tasks they successfully passed. This means that understanding your model is vital; teams must look beyond benchmark scores and evaluate a model's unique defect profile to make informed choices.

Third, these defects are best understood as features of the current technology, not bugs. They appear to be inherent consequences of a methodology that relies on replicating statistical patterns rather than performing semantic analysis. A prime example of this inherent risk is the use of outdated dependencies. Because LLMs are trained on older code, they frequently generate solutions with deprecated APIs or libraries containing known vulnerabilities (CVEs). This underscores the necessity of complementing static analysis with Software Composition Analysis (SCA) to manage a risk vector intrinsic to AI-assisted development.

In conclusion, the integration of LLMs into software development is transformative; however, leveraging this capability effectively calls for informed vigilance. By understanding the potential pitfalls of LLM-generated code and by employing automated analysis tools, the software development community can better navigate this new frontier, taking full advantage of AI while upholding established principles of high-quality, secure, and maintainable software.

\textbf{Future Work.} This study suggests several potential avenues for future investigation:
\begin{itemize}
    \item The impact of various prompt engineering and fine-tuning strategies aimed at mitigating the identified weaknesses in LLM-generated code.
    \item Longitudinal studies tracking the maintainability and technical debt accumulation of software systems with significant LLM contributions.
    \item The effectiveness of LLMs in autonomously refactoring issues identified by static analysis tools, potentially creating a feedback loop for automated code improvement.
    \item Comparative analyses of LLM performance on different programming languages to investigate whether these weaknesses are universal or vary by language.
    \item Research into LLM architectures or training methodologies that could more directly address the challenges LLMs appear to face in generating robust and secure code.
\end{itemize}
Such advancements could contribute to a new generation of LLMs that function as more reliable software engineering assistants.

\newpage
\appendix
\section*{Appendix}
\addcontentsline{toc}{section}{Appendix}

\subsection*{A.1 Bugs categories}
\addcontentsline{toc}{subsection}{A.1 Bugs categories}

\begin{center}
    \captionof{table}{Bugs Categories and Sonar Rules}
    \begin{adjustbox}{width=\textwidth, totalheight=0.85\textheight, keepaspectratio}
        \begin{tabularx}{1.2\textwidth}{l P{5cm} >{\raggedright\arraybackslash}X} 
            \toprule
            \textbf{Bugs Categories} & \textbf{Description / Rule Examples} & \textbf{Sonar Rules Assigned} \\
            \midrule
            API contract violation & Returns, parameters, types, result ignored, equals/hashcode mismatch, contract breaking, incompatible & \texttt{java:S1206 java:S1221 java:S2109 java:S2159 java:S2177 java:S2225 java:S4143 java:S6863 java:S7184 java:S899} \\
            \addlinespace
            Control-flow mistake & Always true/false, conditional blocks, unreachable, logic error, infinite/incorrect branching & \texttt{java:S2189 java:S2583 java:S3923 javabugs:S2190} \\
            \addlinespace
            Exception handling & Swallowing, suppressing, not propagating, mishandling exceptions and finally & \texttt{java:S1143 java:S1163 java:S2142 java:S3551} \\
            \addlinespace
            Resource mgmt/leak & Unclosed resources, leaks, missed close, incorrect freeing & \texttt{java:S2095 java:S2116 java:S2886 java:S5164} \\
            \addlinespace
            Concurrency/threading & Broken synchronization, volatile misuse, thread-unsafe ops, double-checked locking, param sync & \texttt{java:S2168 java:S2222 java:S2445 java:S3077 java:S3078} \\
            \addlinespace
            Type-safety/casts & Broken/illegal/unsafe casts, illegal type use, invalid class cast, use-after-cast, wrong generic & \texttt{java:S1872 java:S2175 java:S2184 java:S2677 javabugs:S6320} \\
            \addlinespace
            Pattern/regex & Broken patterns, regex syntax or misuse, matches empty, ambiguous/redundant expressions & \texttt{java:S2639 java:S5842 java:S5850 java:S5855 java:S5856} \\
            \addlinespace
            Null/data-value & NPE, returns null, invalid/misused Optionals, misused null checks, redundant null-substitute & \texttt{java:S2259 java:S2789 java:S3655} \\
            \addlinespace
            Serialization/Serializable & Missing implements Serializable, writing unserializable object, incompatible object serialization & \texttt{java:S2118 java:S2441} \\
            \addlinespace
            Performance/structure & Stack overflows, repeated computation, non-terminating recursion, infinite loop & \texttt{java:S1751 java:S2119 java:S5998} \\
            \addlinespace
            Data structure & Wrong collection usage, wrong element access, improper generic usage, index-out-of-bounds, assignment & \texttt{javabugs:S6466} \\
            \addlinespace
            Other & Any bug not above & \\
            \bottomrule
        \end{tabularx}
    \end{adjustbox}
\end{center}

\newpage
\subsection*{A.2 Code smells categories}
\addcontentsline{toc}{subsection}{A.2 Code smells categories}

\begin{center}
    \captionof{table}{Code Smells Categories and Sonar Rules}
    \begin{adjustbox}{width=\textwidth, totalheight=0.85\textheight, keepaspectratio}
        \begin{tabularx}{1.15\textwidth}{l P{5cm} >{\raggedright\arraybackslash}X} 
            \toprule
            \textbf{Code Smells Categories} & \textbf{Description / Rule Examples} & \textbf{Rules Assigned} \\
            \midrule
            Naming/style/and documentation & Field/class/interface naming/convention/missing/incorrect/deprecated docs/visibility/badly named constants & \texttt{java:S101 java:S1123 java:S1124 java:S1133 java:S114 java:S115 java:S116 java:S1170 java:S119 java:S1214 java:S1215 java:S1444 java:S2176 java:S3008 java:S6126} \\
            \addlinespace
            Dead/unused/redundant code & Unused variable/method/class/empty class/unused or redundant constants/dead logic/duplicate field/method & \texttt{java:S1126 java:S1130 java:S1144 java:S135 java:S1481 java:S1488 java:S1602 java:S1948 java:S2094 java:S2440 java:S2447 java:S2737 java:S2975 java:S3358 java:S3400 java:S3626 java:S3878 java:S3985 java:S4087 java:S4165 java:S4276 java:S5854} \\
            \addlinespace
            Cognitive/computational complexity & Too complex/method/class size/cyclomatic/brain method/excessive breaks/too many params & \texttt{java:S107 java:S3516 java:S3776 java:S6541} \\
            \addlinespace
            Structure/architecture/layer cycle & Violating package/schematic/dependency/architecture rules or package cycles & \texttt{javaarchitecture:S7027, java:S6809, java:S6833} \\
            \addlinespace
            Collection/generics/param/type & Use generics/types/param types/rawtypes/method ref/lambda/field hiding/diamond op/refactor generics & \texttt{java:S1104 java:S1161 java:S1905 java:S2293 java:S2326 java:S3252 java:S3740 java:S4977 java:S6201 java:S6204} \\
            \addlinespace
            Assignment/field/scope/visibility & Use local not field/make field final/field hiding/static ref/move variable/logic/redundant assignment/visibility & \texttt{java:S127 java:S1854 java:S1450 java:S6213 java:S1117 java:S2209 java:S2201 java:S6837 java:S1994 java:S1193 java:S1611 java:S1141 java:S5993 java:S1165 java:S3305 java:S2147} \\
            \addlinespace
            Control/conditional logic smell & Always true/false/merge if/redundant switch/repeated/complex ternary/duplicate branches & \texttt{java:S1066 java:S1125 java:S1301 java:S1871 java:S2589 java:S4144 java:S5411 java:S6208} \\
            \addlinespace
            Regex/pattern/string/format & Inefficient/bad regex/string concatenation/builder/toString/charsets/formatting/name/misuse of equalsIgnoreCase & \texttt{java:S1149 java:S1153 java:S1192 java:S1210 java:S2629 java:S3457 java:S4635 java:S4719 java:S4738 java:S4968 java:S4973 java:S5413 java:S5843 java:S5866 java:S5869 java:S6019 java:S6035} \\
            \addlinespace
            Deprecated/obsolete APIs & Use of deprecated API/field/missing removal/overdue deprecated code & \texttt{java:S1874 java:S5738 java:S6355} \\
            \addlinespace
            Design \& Framework Best-Practices & Singleton checks/@Autowired advice/codegen/anti-pattern/miscellaneous maintainability & \texttt{java:S112 java:S1155 java:S1168 java:S1612 java:S2139 java:S6548 java:S6829 java:S6833} \\
            \addlinespace
            Other & Any Code Smell not above & \\
            \bottomrule
        \end{tabularx}
    \end{adjustbox}
\end{center}

\newpage
\subsection*{A.3 Vulnerability categories}
\addcontentsline{toc}{subsection}{A.3 Vulnerability categories}

\begin{center}
    \captionof{table}{Vulnerability Categories and Sonar Rules}
    \begin{adjustbox}{width=\textwidth, totalheight=0.85\textheight, keepaspectratio}
        \begin{tabularx}{1.2\textwidth}{l P{5cm} >{\raggedright\arraybackslash}X} 
            \toprule
            \textbf{Vulnerabilities Categories} & \textbf{Description / Rule Examples} & \textbf{Sonar Rules Assigned} \\
            \midrule
            Hard-coded credentials & Hardcoded passwords/keys/secrets found in code & \texttt{java:S6437} \\
            \addlinespace
            Path-traversal \& injection & Path traversal, unsafe archive/file/URL constructs, user-controlled path/file/cookie/URL/JSON & \texttt{java:S6377 javasecurity:S2083 javasecurity:S5144 javasecurity:S6096 javasecurity:S6287 javasecurity:S6549} \\
            \addlinespace
            Cryptography misconfiguration & Weak/incorrect cipher/IV, insecure encryption, insecure hash, improper cipher mode/padding & \texttt{java:S2053 java:S3329 java:S5445 java:S5542 java:S5547} \\
            \addlinespace
            Certificate-validation omissions & Missing server hostname/certificate validation in SSL/TLS & \texttt{java:S4830 java:S5527} \\
            \addlinespace
            JWT signature not verified & JWT signature is not checked/verified before being used & \texttt{java:S5659} \\
            \addlinespace
            XML External Entity (XXE) & External entity expansion, XML parser not protected from external entities & \texttt{java:S2755} \\
            \addlinespace
            JSON-injection risk & Direct construction of JSON from untrusted/user-controlled data & \texttt{javasecurity:S6398} \\
            \addlinespace
            Inadequate error handling (I/O) & Failing to catch, propagate, or properly handle critical exceptions from untrusted sources & \texttt{java:S1989} \\
            \addlinespace
            Other & Any vulnerability not above & \\
            \bottomrule
        \end{tabularx}
    \end{adjustbox}
\end{center}

\newpage

\bibliographystyle{ieeetr}

\bibliography{\jobname}

\end{document}